\documentstyle{lamuphys}                                               
\begin{document}
\title{Making manifest the symmetry enhancement for coincident BPS branes}

\author{Sergei\, V. Ketov}

\institute{Institut f\"ur Theoretische Physik, Universit\"at Hannover, Appelstra{\ss}e 2,
D-30167 Hannover, Germany}

\maketitle

\begin{abstract}
We consider $g$ {\it coincident} M-5-branes on top of each other, in the KK monopole background 
$Q$ of multiplicity $N$. The worldvolume of each M-5-brane is supposed to be given by the local 
product of the four-dimensional spacetime $R^{1,3}$ and an elliptic curve. In the coincidence 
limit, all these curves yield a single (Seiberg-Witten) hyperelliptic curve $\Sigma_g$, while the 
gauge symmetry is enhanced to $U(N)$. We make this gauge symmetry enhancement manifest by 
considering the {\it hypermultiplet} LEEA which is given by the spacetime N=2 non-linear 
sigma-model (NLSM) having $Q$ as the target space. The hyper-K\"ahler manifold $Q$ is given by the 
multicentre Taub-NUT space, which in the coincidence limit amounts to the multiple Eguchi-Hanson 
(ALE) space $Q_{\rm mEH}$. The NLSM is most naturally described in terms of the hyper-K\"ahler coset 
construction on $SU(N,N)/U(N)$ in harmonic superspace, by using the auxiliary (in classical theory) 
N=2 vector superfields as Lagrange multipliers, with FI terms resolving the $A_{N}$ singularity. 
The Maldacena limit, in which the hypermultiplet LEEA becomes extended to the N=4 SYM with the gauge
group $U(N)$, arises in quantum field theory due to a dynamical generation of the N=2 vector 
and hypermultipet superfields, when sending the FI terms to zero.
\end{abstract}



\let\under=\unt                 
\let\ced=\ce                    
\let\du=\du                     
\let\um=\Hu                     
\let\sll=\lp                    
\let\Sll=\Lp                    
\let\slo=\os                    
\let\Slo=\Os                    
\let\tie=\ta                    
\let\br=\ub                     


\def\a{\alpha}
\def\b{\beta}
\def\c{\chi}
\def\d{\delta}
\def\e{\epsilon}
\def\f{\phi}
\def\g{\gamma}
\def\h{\eta}
\def\i{\iota}
\def\j{\psi}
\def\k{\kappa}
\def\l{\lambda}
\def\m{\mu}
\def\n{\nu}
\def\o{\omega}
\def\p{\pi}
\def\q{\theta}
\def\r{\rho}
\def\s{\sigma}
\def\t{\tau}
\def\u{\upsilon}
\def\x{\xi}
\def\z{\zeta}
\def\D{\Delta}
\def\F{\Phi}
\def\G{\Gamma}
\def\J{\Psi}
\def\L{\Lambda}
\def\O{\Omega}
\def\P{\Pi}
\def\Q{\Theta}
\def\S{\Sigma}
\def\U{\Upsilon}
\def\X{\Xi}


\def\ve{\varepsilon}
\def\vf{\varphi}
\def\vr{\varrho}
\def\vs{\varsigma}
\def\vq{\vartheta}


\def\ca{{\cal A}}
\def\cb{{\cal B}}
\def\cc{{\cal C}}
\def\cd{{\cal D}}
\def\ce{{\cal E}}
\def\cf{{\cal F}}
\def\cg{{\cal G}}
\def\ch{{\cal H}}
\def\ci{{\cal I}}
\def\cj{{\cal J}}
\def\ck{{\cal K}}
\def\cl{{\cal L}}
\def\cm{{\cal M}}
\def\cn{{\cal N}}
\def\co{{\cal O}}
\def\cp{{\cal P}}
\def\cq{{\cal Q}}
\def\car{{\cal R}}
\def\cs{{\cal S}}
\def\ct{{\cal T}}
\def\cu{{\cal U}}
\def\cv{{\cal V}}
\def\cw{{\cal W}}
\def\cx{{\cal X}}
\def\cy{{\cal Y}}
\def\cz{{\cal Z}}


\def\Sc#1{{\hbox{\sc #1}}}      
\def\Sf#1{{\hbox{\sf #1}}}      



\def\slpa{\slash{\pa}}                            
\def\slin{\SLLash{\in}}                                   
\def\bo{{\raise-.5ex\hbox{\large$\Box$}}}               
\def\cbo{\Sc [}                                         
\def\pa{\partial}                                       
\def\de{\nabla}                                         
\def\dell{\bigtriangledown}                             
\def\su{\sum}                                           
\def\pr{\prod}                                          
\def\iff{\leftrightarrow}                               
\def\conj{{\hbox{\large *}}}                            
\def\ltap{\raisebox{-.4ex}{\rlap{$\sim$}} \raisebox{.4ex}{$<$}}   
\def\gtap{\raisebox{-.4ex}{\rlap{$\sim$}} \raisebox{.4ex}{$>$}}   
\def\TH{{\raise.2ex\hbox{$\displaystyle \bigodot$}\mskip-4.7mu \llap H \;}}
\def\face{{\raise.2ex\hbox{$\displaystyle \bigodot$}\mskip-2.2mu \llap {$\ddot
        \smile$}}}                                      
\def\dg{\sp\dagger}                                     
\def\ddg{\sp\ddagger}                                   

\font\tenex=cmex10 scaled 1200


\def\sp#1{{}^{#1}}                              
\def\sb#1{{}_{#1}}                              
\def\oldsl#1{\rlap/#1}                          
\def\slash#1{\rlap{\hbox{$\mskip 1 mu /$}}#1}      
\def\Slash#1{\rlap{\hbox{$\mskip 3 mu /$}}#1}      
\def\SLash#1{\rlap{\hbox{$\mskip 4.5 mu /$}}#1}    
\def\SLLash#1{\rlap{\hbox{$\mskip 6 mu /$}}#1}      
\def\PMMM#1{\rlap{\hbox{$\mskip 2 mu | $}}#1}   %
\def\PMM#1{\rlap{\hbox{$\mskip 4 mu ~ \mid $}}#1}       %
\def\Tilde#1{\widetilde{#1}}                    
\def\Hat#1{\widehat{#1}}                        
\def\Bar#1{\overline{#1}}                       
\def\sbar#1{\stackrel{*}{\Bar{#1}}}             
\def\bra#1{\left\langle #1\right|}              
\def\ket#1{\left| #1\right\rangle}              
\def\VEV#1{\left\langle #1\right\rangle}        
\def\abs#1{\left| #1\right|}                    
\def\leftrightarrowfill{$\mathsurround=0pt \mathord\leftarrow \mkern-6mu
        \cleaders\hbox{$\mkern-2mu \mathord- \mkern-2mu$}\hfill
        \mkern-6mu \mathord\rightarrow$}
\def\dvec#1{\vbox{\ialign{##\crcr
        \leftrightarrowfill\crcr\noalign{\kern-1pt\nointerlineskip}
        $\hfil\displaystyle{#1}\hfil$\crcr}}}           
\def\dt#1{{\buildrel {\hbox{\LARGE .}} \over {#1}}}     
\def\dtt#1{{\buildrel \bullet \over {#1}}}              
\def\der#1{{\pa \over \pa {#1}}}                
\def\fder#1{{\d \over \d {#1}}}                 


\def\frac#1#2{{\textstyle{#1\over\vphantom2\smash{\raise.20ex
        \hbox{$\scriptstyle{#2}$}}}}}                   
\def\half{\frac12}                                        
\def\sfrac#1#2{{\vphantom1\smash{\lower.5ex\hbox{\small$#1$}}\over
        \vphantom1\smash{\raise.4ex\hbox{\small$#2$}}}} 
\def\bfrac#1#2{{\vphantom1\smash{\lower.5ex\hbox{$#1$}}\over
        \vphantom1\smash{\raise.3ex\hbox{$#2$}}}}       
\def\afrac#1#2{{\vphantom1\smash{\lower.5ex\hbox{$#1$}}\over#2}}    
\def\partder#1#2{{\partial #1\over\partial #2}}   
\def\parvar#1#2{{\d #1\over \d #2}}               
\def\secder#1#2#3{{\partial^2 #1\over\partial #2 \partial #3}}  
\def\on#1#2{\mathop{\null#2}\limits^{#1}}               
\def\bvec#1{\on\leftarrow{#1}}                  
\def\oover#1{\on\circ{#1}}                              

\def\[{\lfloor{\hskip 0.35pt}\!\!\!\lceil}
\def\]{\rfloor{\hskip 0.35pt}\!\!\!\rceil}
\def\Lag{{\cal L}}
\def\du#1#2{_{#1}{}^{#2}}
\def\ud#1#2{^{#1}{}_{#2}}
\def\dud#1#2#3{_{#1}{}^{#2}{}_{#3}}
\def\udu#1#2#3{^{#1}{}_{#2}{}^{#3}}
\def\calD{{\cal D}}
\def\calM{{\cal M}}

\def\szet{{${\scriptstyle \b}$}}
\def\ulA{{\un A}}
\def\ulM{{\underline M}}
\def\cdm{{\Sc D}_{--}}
\def\cdp{{\Sc D}_{++}}
\def\vTheta{\check\Theta}
\def\fracm#1#2{\hbox{\large{${\frac{{#1}}{{#2}}}$}}}
\def\ha{{\fracmm12}}
\def\tr{{\rm tr}}
\def\Tr{{\rm Tr}}
\def\itrema{$\ddot{\scriptstyle 1}$}
\def\ula{{\underline a}} \def\ulb{{\underline b}} \def\ulc{{\underline c}}
\def\uld{{\underline d}} \def\ule{{\underline e}} \def\ulf{{\underline f}}
\def\ulg{{\underline g}}
\def\items#1{\\ \item{[#1]}}
\def\ul{\underline}
\def\un{\underline}
\def\fracmm#1#2{{{#1}\over{#2}}}
\def\footnotew#1{\footnote{\hsize=6.5in {#1}}}
\def\low#1{{\raise -3pt\hbox{${\hskip 0.75pt}\!_{#1}$}}}

\def\Dot#1{\buildrel{_{_{\hskip 0.01in}\bullet}}\over{#1}}
\def\dt#1{\Dot{#1}}
\def\DDot#1{\buildrel{_{_{\hskip 0.01in}\bullet\bullet}}\over{#1}}
\def\ddt#1{\DDot{#1}}

\def\Tilde#1{{\widetilde{#1}}\hskip 0.015in}
\def\Hat#1{\widehat{#1}}


\newskip\humongous \humongous=0pt plus 1000pt minus 1000pt
\def\caja{\mathsurround=0pt}
\def\eqalign#1{\,\vcenter{\openup2\jot \caja
        \ialign{\strut \hfil$\displaystyle{##}$&$
        \displaystyle{{}##}$\hfil\crcr#1\crcr}}\,}
\newif\ifdtup
\def\panorama{\global\dtuptrue \openup2\jot \caja
        \everycr{\noalign{\ifdtup \global\dtupfalse
        \vskip-\lineskiplimit \vskip\normallineskiplimit
        \else \penalty\interdisplaylinepenalty \fi}}}
\def\li#1{\panorama \tabskip=\humongous                         
        \halign to\displaywidth{\hfil$\displaystyle{##}$
        \tabskip=0pt&$\displaystyle{{}##}$\hfil
        \tabskip=\humongous&\llap{$##$}\tabskip=0pt
        \crcr#1\crcr}}
\def\eqalignnotwo#1{\panorama \tabskip=\humongous
        \halign to\displaywidth{\hfil$\displaystyle{##}$
        \tabskip=0pt&$\displaystyle{{}##}$
        \tabskip=0pt&$\displaystyle{{}##}$\hfil
        \tabskip=\humongous&\llap{$##$}\tabskip=0pt
        \crcr#1\crcr}}


\section{Brane technology and KK monopoles}

The (Seiberg-Witten-type) exact solution to the N=2 super-QCD can be identified with the LEEA 
of the effective (called N=2 MQCD) field theory defined in the worldvolume of the {\it single} 
M-5-brane, given by the local product of the uncompacftified four-dimensional spacetime 
$R^{1,3}$ and the hyperelliptic curve $\S_g$ of genus $g=N-1$: \cite{witten} 
(see, e.g. \cite{ber,m1,m2} for a review or an introduction). The hyperelliptic curve 
$\S_g$ is supposed to be holomorphically embedded into the hyper-K\"ahler four-dimensional multiple
Taub-NUT space $Q_{\rm mTN}$ associated with the multiple KK-monopole. The identification of the 
{\it low-energy effective actions} (LEEA) in these two, apparently very different, field theories 
(the N=2 super-QCD in the Coulomb branch, on the one side, and the N=2 MQCD defined in the 
M-5-brane worldvolume, on the other side) is highly non-trivial, since the former is defined as the 
leading contribution to the {\it quantum} LEEA in a gauge field theory, whereas the latter is 
determined by the {\it classical} M-5-brane dynamics or the $D=11$ supergravity equations of motion, 
whose BPS solutions preserving some part of supersymmetry are the M-theory branes under 
consideration.

\subsection{Multiple KK monopole}

The multiple KK monopole is a non-singular BPS solution to the eleven-dimensional 
supergravity equations of motion, given by the product of the seven-dimensional (flat) 
Minkowski spacetime $R^{1,6}$ and the four-dimensional Euclidean multicentre Taub-NUT 
space $Q_{\rm mTN}$ (\cite{town}):
\begin{eqnarray}
ds^2_{[11]} & =& dx^{\m}dx^{\n}\h_{\m\n}+H(d\vec{y})^2+H^{-1}(d\varrho+\vec{C}\cdot
d\vec{y})^2 \enspace,  \nonumber \label{met} \\
\vec{\de}\times\vec{C}& =& \vec{\de}H~,\qquad F_{(4)}\equiv dA_{(3)}=0 \enspace  ,
\end{eqnarray}
where $\m=0,1,2,3,7,8,9$, $\vec{y}=\{y_i\}$, $i=4,5,6$, the eleventh coordinate $\varrho$ is 
supposed to be periodic (with the period $2\p k$ -- this is just necessary to avoid conical
singularities of the metric), while the harmonic function $H(\vec{y})$ is given by
\begin{equation} 
H(\vec{y})=1 +\sum_{p=1}^{N}\fracmm{\abs{k}}{2\abs{\vec{y}-\vec{y}_p}}\enspace .\label{harm}
\end{equation}
The moduli $(k,\vec{y}_i)$ can be interpreted as (equal) charges and the locations
of the KK monopoles, respectively. The multiple Taub-NUT space can be thought of as a 
non-trivial bundle (Hopf fibration) with the base $R^3$ and the fiber $S^1$ of magnetic charge
$k$. There exist $N$ linearly independent normalizable self-dual harmonic 2-forms $\o_p$ in 
$Q_{\rm mTN}$, which satisfy the orthogonality condition (see \cite{gorv}, and references 
therein)
\begin{equation}
\fracmm{1}{(2\p k)^2}\int_{Q_{\rm mTN}} \,\o_p\wedge \o_q=\d_{pq} \enspace .\label{ortho}
\end{equation}
As is clear from eq.~(\ref{met}), two adjacent KK monopoles are connected by a homology 2-sphere 
having poles at the positions of the two monopoles. Near a singularity of $H$, the KK circle $S^1$ 
contracts to a point. A {\it holomorphic} embedding of the Seiberg-Witten spectral curve $\S$ into
the hyper-K\"ahler manifold $Q_{\rm mTN}$ is the consequence of the {\it BPS condition} 
(\cite{mikh}) 
\begin{equation}
{\rm Area}\low{\S}=\abs{\int_{\S}\,\O_{\S} } \enspace  ,\label{bpsm}
\end{equation}
where $\O_{\S}$ is the pullback of the K\"ahler $(1,1)$ form $\O$ of $Q_{\rm mTN}$ on $\S$. 
Any four-dimensional hyper-K\"ahler manifold, in fact, possesses a holomorphic $(2,0)$ form $\o$, 
which is simply related to the K\"ahler form $\O$ as
\begin{equation}
\O^2=\o\wedge \bar{\o} \enspace  .\label{hyk}
\end{equation}
The BPS states in M-theory, whose zero modes appear in the effective field theory defined in
the M-5-brane worldvolume, correspond to the minimal area (BPS) M-2-branes ending on the M-5-brane.
The {\it spacial} topology of an M-2-brane determines the type of the corresponding supermultiplet 
in the spacetime: a cylinder $(Y)$ leads to an N=2 vector multiplet, whereas a disc $(D)$ gives rise
to a hypermultiplet (\cite{mikh}). Since the pullback $\o_Y$ on $Y$ is closed (\cite{heyi}), there
exists a meromorphic differential $\l_{SW}$ satisfying the relations $\o_Y=d\l_{SW}$ and
\begin{equation}
Z=\int_{Y}\,\o_Y =\oint_{\pa Y}\l_{\rm SW} \enspace  ,\label{cc}
\end{equation}
where $Z$ is the central charge and $\pa Y\in \S$. Hence, $\l_{\rm SW}$ can be identified with the 
Seiberg-Witten differential, which determines the spacetime N=2 gauge LEEA (see also \cite{fsp}).

\subsection{N=2 QCD LEEA in Coulomb branch from brane dynamics} 

The geometrical origin and the physical interpretation of the hyperelliptic curve $\S_g$, 
parameterizing the exact Seiberg-Witten solution to the LEEA of $N=2$ supersymmetric QCD 
in the Coulomb branch, becomes transparent when using 
brane technology after M-theory resolution of UV singularities (\cite{witten}), where $\S_g$ 
appears to be the part of the M-5-brane worldvolume in eleven dimensions. The Nambu-Goto (NG) 
term (proportional to the M-5-brane worldvolume) of the effective M-5-brane action in the 
low-energy approximation gives rise to the scalar {\it non-linear 
sigma model} (NLSM) having the special geometry after the KK compactification of the 
six-dimensional NG action on the Seiberg-Witten curve $\S_g$ down to four spacetime dimensions. 
This is enough to unambiguously restore the full $N=2$ supersymmetric Seiberg-Witten LEEA, 
by the use of $N=2$ supersymmetrization of the special bosonic NLSM, when considering its complex 
scalars as the leading components of abelian $N=2$ vector multiplets in four spacetime 
dimensions and then deducing the Seiberg-Witten holomorphic potential out of the already 
derived special K\"ahler NLSM potential (see \cite{m2} for a review).
 
Being applied to a derivation of the {\it hypermultiplet} LEEA of $N=2$ super-QCD in the 
Coulomb branch, brane technology suggests to dimensionally reduce the effective action of
a D-6-brane (to be described in M-theory by a KK-monopole) down to four spacetime dimensions
(\cite{m2}). In a static gauge for the D-6-brane, the indiced metric in the brane worldvolume 
is given by
\begin{equation}
g_{\m\n}=\h_{\m\n}+G_{mn}\pa_{\m}y^m\pa_{\n}y^n \enspace ,\label{induced}
\end{equation}
where $\m,\n=0,1,2,3,7,8,9$, $m,n=4,5,6,10$, and $G_{mn}$ is the multicentre ETN metric. After
expanding the NG-part of the D-6-brane effective action  
\begin{equation}
S_{\rm NG} =\int d^7\x\,\sqrt{-\det(g_{\m\n})} \label{NG}
\end{equation}
up to the second-order in the spacetime derivatives, and performing plain dimensional reduction
down to four dimensions, one arrives at the hyper-K\"ahler NLSM
\begin{equation}
S[y]=-\ha \int d^4x\,G_{mn}(y)\pa_{\un{\m}}y^m\pa^{\un{\m}}y^n~,\qquad
\un{\m}=0,1,2,3 \enspace ,\label{hyper}
\end{equation}
whose N=2 supersymmetrization yields the full hypermultiplet LEEA, in precise agreement
with the N=2 supersymmetric quantum field theory calculations in harmonic superspace 
(\cite{ikz}).

\section{Symmetry enhancement: two coincident D-6-branes}

As is well known, the isolated singularities of the harmonic function (\ref{harm}) are just 
the coordinate singularities of the {\it eleven}-dimensional metric (\ref{met}), though they 
are truly singular with respect to the (dimensionally reduced) {\it ten}-dimensional metric to
be associated with the D-6-branes in the type-IIA picture. The physical significance of these
ten-dimensional singularities can therefore be understood due to the illegitimate neglect of 
the KK modes related to the compactification circle $S^1$ in ten dimensions, since the KK 
particles (also called D-0-branes) become massless near the D-6-brane core (\cite{town}). Their
inclusion is equivalent to accounting the instanton corrections in the four-dimensional N=2 
supersymmetric gauge field theory. 

When some parallel and similarly oriented D-branes coincide, it is usually accompanied by 
{\it symmetry enhancement} (\cite{ht},\cite{witten2}). Since the brane singularities become 
non-isolated in the coincidence limit, first, they have to be resolved by considering the branes to
be separated by some distance $r$. Then one takes the limit $r\to 0$. In the simplest non-trivial 
case of two D-6-branes, one can consider the harmonic function 
\begin{equation}
H(\vec{y})=\l+\fracmm{1}{2}\left\{ \fracmm{1}{\abs{\vec{y}-\x\vec{e}}}
+ \fracmm{1}{\abs{\vec{y}+\x\vec{e}}}\right\}~,\qquad \quad r=2\x \enspace ,
\label{twob}
\end{equation}
describing the double-centered Taub-NUT metric in (\ref{met}) with a non-vanishing constant 
potential at infinity, whose both centers are on the line $\vec{e}$ in sixth direction, 
$\vec{e}^2=1$. After being substituted into eq.~(\ref{met}), eq.~(\ref{twob}) describes two
parallel and similarly oriented KK-monopoles in M-theory, which dimensionally reduce to the
double D-6-brane configuration in ten dimensions. The homology 2-sphere connecting the KK 
monopoles contracts to a point in this limit, which gives rise to a curvature singularity of 
the dimensionally reduced metric in ten dimensions. From the eleven-dimensional viewpoint, 
M-2-branes can wrap about the 2-sphere connecting the KK monopoles, while the energy of the 
wrapped M-2-brane is proportional to the area of the sphere (\cite{ht}). When the sphere collapses, 
its area vanishes and, hence, an additional massless vector state appears due to the zero mode of 
the wrapped M-2-brane. One thus expects the gauge symmetry enhancement from  $U(1)\times U(1)$ to 
$U(2)$ assiciated with the $A_1$-type singularity (\cite{ov}). From the ten-dimensional viewpoint, 
the wrapped M-2-branes are just the (6-6) superstrings stretched between the D-6-branes, so that it 
is the massless zero modes of these 6-6 superstrings that become massless in the coincidence limit.

In order to make this symmetry enhancement manifest, let's start from the hypermultiplet low-energy 
effective action, which is obtained by spacetime N=2 supersymmetrization of the bosonic NLSM 
(\ref{hyper}) and whose hyper-K\"ahler (double Taub-NUT) metric is determined by the harmonic 
function (\ref{twob}). In terms of this NLSM metric, the symmetry enhancement amounts to 
the appearance of gauged isometries in the NLSM target space, while the latter can be made manifest 
in the N=2 harmonic superspace, as we are now going to demonstrate. First, let's note that the N=2 
supersymmetric double Taub-NUT NLSM is known to be equivalent to the one with the {\it mixed} 
(=Eguchi-Hanson-Taub-NUT) metric (\cite{gorv}). The mixed NLSM is described by the following N=2 
harmonic superfield action over the analytic subspace:
\begin{eqnarray}
S_{\rm mixed}[q,V]  =  \int_{\rm analytic}\,& & \left\{ q^{aA+}D^{++}_Z q^+_{aA}+V_3^{++}
\left(\frac{1}{2}\ve^{AB3}q^{a+}_A q^+_{aB}+\x^{++}\right)\right. \enspace  \nonumber \\
& &\left. + \frac{1}{4}\l(q^{aA+}q^+_{aA})^2\right\} \enspace  ,\label{haction} 
\end{eqnarray}
whose analytic hypermultiplet superfields $q^+_A$, $A=1,2$, in the pseudo-real notation 
$q^{a+}=(\sbar{q}{}^+,q^+)$, $a=1,2$, belong to a linear representation $\underline{2}$ of 
$SU(1,1)$ whose $U(1)$ subgroup is gauged by the use of the auxiliary N=2 vector gauge analytic 
superfield  (Lagrange multiplier) $V_3^{++}$. The parameters $\l$ in 
eqs.~(\ref{twob}) and (\ref{haction}) can be identified, whereas the parameter 
$\x^{++}=\x^{ij}u^+_iu^+_j$ in eq.~(\ref{haction}) is simply related to the constant $\x$ appearing 
in eq.~(\ref{twob}) as $\abs{\x}{}^2=-\frac{1}{8}(\x^{ij})^2$. The hyper-K\"ahler NLSM metric, 
which is 
deduced out of the superspace action (\ref{haction}) after eliminating all the auxiliary fields in 
components, yields the double Taub-NUT metric, as can be verified by explicit calculation  
(\cite{gorv}). This is, in fact, ensured by the manifest $U(1)_A\times U(1)_{\rm PG}$ symmetry 
of the superspace action 
(\ref{haction}), where the first $U(1)_A$ factor is the unbroken part of the $SU(2)_A$ 
automorphisms of the $N=2$ supersymmetry algebra rotating two supercharges, whereas the second 
$U(1)_{\rm PG}$ symmetry only acts on the pseudo-real indices $a=1,2$ and thus implies an abelian 
isometry of the NLSM metric. In fact, any four-dimensional hyper-K\"ahler metric having the 
$U(1)_{\rm PG}$ isometry is a multicentre Taub-NUT metric (see \cite{gorv} and references therein). 
The mixed four-dimensional hyper-K\"ahler metric of the N=2 supersymmetric NLSM (\ref{haction}) 
clearly interpolates between the {\it Eguchi-Hanson} (EH) metric $(\l=0)$ and the Taub-NUT $(\x=0)$,
both having the maximal isometry group $U(2)$. The action of the $U(2)$ isometry is linear in both 
limiting cases, while it is even holomorphic in the second case. In the harmonic superspace 
approach, this symmetry enhancement can be simply understood either as the restoration of the 
$SU(2)_A$ automorphism symmetry in the Taub-NUT limit, or as the restoration of the $SU(2)_{\rm PG}$
symmetry in the Eguchi-Hanson limit.

When using the component results collected in the Appendix B of (\cite{m2}), it is not difficult
to verify that the spacetime vector gauge field belonging to the $V^{++}$ supermultiplet becomes 
{\it dynamical} due to quantum fluctuations of the hypermultiplets $q^+_A$ (see, e.g. sect.~8.3 of 
\cite{pol} for a similar phenomenon in two spacetime dimensions). After taking into account the 
$U(1)$ gauge symmetry and N=2 supersymmetry, this implies the dynamical generation of the extra 
physical $N=2$ vector gauge superfield, in full agreement with the predictions of brane technology.

\section{Symmetry enhancement: $N$ coincident D-6-branes}

The $D=11$ supergravity approximation to M-theory is only valid for well-separated KK monopoles.
When the KK monopoles coincide, their low-energy dynamics is to be approximated by weakly coupled 
superstrings propagating in the multi-Eguchi-Hanson (ALE) background (\cite{sen}). This background 
naturally originates from the multi-Taub-NUT space. Indeed, if all the D-6-branes coincide, they 
can be described in M-theory by sending all the moduli $\vec{y}_p$ in the harmonic function 
(\ref{harm}) to zero, so that the additive constant (asymptotic potential), which is equal to one in
eq.~(\ref{harm}), can be ignored near the core of the D-6-branes on top of each other. The 
multi-Eguchi-Hanson space thus possesses $A_{N-1}$ simple singularity which implies the enhanced 
non-abelian gauge symmetry $SU(N)$ in the effective supersymmetric field theory defined in the 
common worldvolume of the coincident D-6-branes.

The effective gauge field theory is supposed to be defined in the limit where the gravity
decouples. The $D=11$ supergravity has a 3-form $A^{[11]}_{(3)}$ which is decomposed in the 
full spacetime given by the product of the D-6-brane worldvolume $R^{1,6}$ and the 
multi-Taub-NUT space $Q_{\rm mTN}$ as 
\begin{equation}
A^{[11]}_{(3)}=\sum_{p=1}^{N} A_{p(1)}^{[7]}\wedge \o_{p(2)}^{[4]} \enspace  ,\label{decom}
\end{equation}
where the 2-forms $\o_p$ in $Q_{\rm mTN}$ have been introduced in subsect.~1.1, whereas
$A_p$ are $N$ massless vectors (1-forms) in $R^{1,6}$. In addition, there are $3N$ scalar 
fields associated with the translational zero modes (or moduli) $\vec{y}_p$. All these vectors
and scalars are the bosonic components of $N$ massless vector supermultiplets in $1+6$ 
dimensions, each having $8_{\rm B}+8_{\rm F}$ on-shell components. The gauge group of the 
effective field theory (in the case of separated KK monopoles) in the Coulomb branch is
therefore given by $U(1)^N$. Since the intersection matrix of 2-cycles in $Q_{\rm mTN}$ is just
given by the Cartan matrix of $A_{N-1}$, the abelian gauge symmetry $U(1)^{N}$ is to be enhanced 
to $U(N)$ (thus defining the {\it non-abelian} `Coulomb branch') in the coincidence limit 
(\cite{sen}). The area of the 2-cycles vanishes in this limit, so that the M-2-branes wrapped 
around these 2-cycles give rise to the additional massless vectors which are the M-2-brane zero 
modes. In the type-IIA picture, the 6-6 strings stretched between separated D-6-branes do not 
contribute to the effective LEEA in the Coulomb branch. However, since the zero modes of 6-6 strings
become massless when the brane separation vanishes, they do contribute to the LEEA in the 
non-abelian Coulomb branch. After plain dimensional reduction from $R^{1,6}$ to $R^{1,3}$, the 
effective N=1 super-Yang-Mills theory in $1+6$ dimensions yields the N=4 super-Yang-Mills theory in 
$1+3$ dimensions, which has the same number of on-shell components.  

We arrive at the same conclusions on the quantum field theory side of the story in the 
four-dimensional spacetime, when we consider the corresponding hypermultiplet LEEA given by the 
gauged N=2 supersymmetric NLSM associated with the coset $SU(N,N)/U(N)$ in harmonic superspace,  
\begin{equation}
S_{\rm non-abelian}[q,V] = \int_{\rm analytic}\left\{ \tr\low{\rm F}
\left(\sbar{q}{}^{+}\cd^{++} q^+\right)+\tr\low{\rm C}\left(V^{++}\cdot\x^{++} \right) \right\}
\enspace  ,\label{naction}
\end{equation}
where $\cd^{++}=D_Z^{++}+iV^{++}$, the covariant derivative $D_Z^{++}$ has central charge $Z$, 
and the hypermultiplets $q$ are supposed to belong to the fundamental (F) representation of 
$SU(N,N)$ whose $U(N)$ subgroup is gauged by the use of the N=2 vector gauge superfield $V^{++}$ 
valued in the Lie algebra of $U(N)$. The Fayet-Iliopoulos (FI) terms are now valued in the Cartan 
(C) subalgebra of $U(N)$. These FI terms are apparently necessary in the action, in order to 
resolve the $A_{N}$ singularity which would appear in their absence.

The N=2 hypermultiplet superpropagator in the harmonic superspace 
$\cz=\{x^{\m},\q^i_{\a},\bar{\q}_i^{\dt{\a}}\}$ reads (see, e.g., \cite{ikz}) 
\begin{equation}
i\VEV{q^+(1)\sbar{q}{}^+(2)}=-\fracmm{1}{\bo_1^Z}(D^+_1)^4(D^+_2)^4\left\{ \d^{12}(\cz_1-\cz_2)
\fracmm{e^{v_{Z}(2)-v_{Z}(1)} } { (u^+_1u^+_2)^3}\right\}\enspace ,
\label{pro}
\end{equation}
where we have introduced the so-called `bridge' $v_Z$ associated with the abelian N=2 vector gauge 
superfield whose N=2 chiral field strength is constant and equal to the central charge $Z$. It is 
now straightforward to evaluate the local part of the one-loop effective action $i\Tr\log\cd^{++}$ 
in the LEEA approximation, with the latter being defined by the condition $\abs{k}^2\ll\abs{\x}^2$ 
for all external momenta $k^{\m}$. This should yield
\begin{eqnarray}
S_{\rm N=2~SYM}[V]  = & &\fracmm{1}{g^2\low{\rm YM}}\int_{\rm full}\, \tr \sum^{\infty}_{n=2}
\fracmm{(-i)^n}{n}\int du_1du_2\cdots du_n\times  \enspace  \nonumber \\
& & \times \fracmm{ V^{++}(\cz,u_1)V^{++}(\cz,u_2)\cdots V^{++}(\cz,u_n)}{(u^+_1u^+_2)
(u^+_2u^+_3)\cdots (u^+_nu^+_1)} \enspace  ,\label{2sym}
\end{eqnarray}
with the induced gauge coupling $g^{-2}_{\rm YM}\sim N\log\left(1+\fracmm{\abs{\x}}{\abs{Z}^2}
\right)$. The action (\ref{2sym}) is known as the N=2 supersymmetric Yang-Mills action in harmonic 
superspace (\cite{zup}). 

I conclude this section with a few comments. 

If the $(1+6)$-dimensional $N=1$ supersymmetric 
effective field theory were compactified on the circle $S^1$, this would give rise to the gauge 
field theory in $(1+5)$ dimensions, whose T-dual is an $(2,0)$ supersymmetric gauge field theory 
with $N$ tensor multiplets. Therefore, in the type-IIB case, we do not get an enhanced gauge 
symmetry but $N$ tensor multiplets instead. 

A truly different gauge symmetry enhancement pattern 
appears when $N$ of D-6-branes come on top of an {\it orientifold} six-plane, which leads to the 
$SO(2N)$ gauge symmetry (\cite{ov}). In M-theory, the orientifold six-plane can be represented by 
the {\it Atiyah-Hitchin} space (\cite{ati}) instead of a KK monopole (\cite{sen}). Indeed, far from 
the origin the Atiyah-Hitchin space has the topology $R^3\times S^1/\ct_4$, i.e. it looks like 
$Q_{\rm mTN}$ whose points are now supposed to be identified under the action of the discrete 
symmetry $\ct_4$ reversing signs of all four coordinates of $Q_{\rm mTN}$. This matches the 
definition of the orientifold six-plane according to \cite{sen}.

\section{Large $N$ limit}

In order to reproduce the Seiberg-Witten-type solution to $N=2$ super-QCD from M-theory, merely a  
{\it single} and smooth M-5-brane in a KK-monopole background is needed. The M-5-brane worldvolume
is to be compactified on the SW curve $\S_g$ down to $(1+3)$ dimensions i.e. to the worldvolume of 
a {\it D-3-brane}. Taking $N$ M-5-branes (whose worldvolume is given by the local product 
$R^{1,3}\times\S_1$) and allowing them to coincide in the background of $N$ KK monopoles yields 
(at a generic point in the moduli space) an $N=2$ supersymmetric gauge field theory with the 
non-abelian gauge group $U(N)$ as the LEEA in the common (macroscopically $(1+3)$-dimensional) brane
worldvolume. The KK monopoles merge in this process too, which also implies extra massless 
hypermultiplets in the LEEA, thus increasing their number to $N^2$, i.e. to the {\it adjoint} 
representation of the gauge group.

The coincidence limit corresponds to sending $\abs{\x}\to 0$ in the spacetime LEEA which is now 
given by a sum of the gauge-invariant action of a massless hypermultiplet in the adjoint of $U(N)$ 
and the N=2 super-Yang-Mills action (\ref{2sym}). This sum,
\begin{equation}
\int_{\rm analytic} \tr\low{\rm ad}\left(\sbar{q}{}^{+}\cd^{++} q^+\right)+ S_{\rm N=2~SYM}
\equiv S_{\rm N=4~SYM}\enspace  ,\label{4sym}
\end{equation}
is, however, nothing but the N=4 supersymmetric Yang-Mills action in the N=2 harmonic superspace~!
The induced super-Yang-Mills gauge coupling constant in the limit $\abs{\x}\to 0$ is given by
\begin{equation}
Ng^{2}_{\rm YM}\sim \fracmm{\abs{Z}^2}{\abs{\x}}\enspace  ,\label{coup}
\end{equation}
so that the non-vanishing central charge is really necessary for a dynamical generation of the N=2
supersymmetric Yang-Mills fields.

Our result is closely related to the recent conjecture of Maldacena (1997). He discussed `simple' 
M-5-branes, having no Riemann surface in their worldvolumes, in the particular limit (down the 
`throat') given by the product $AdS_7\times S^4$ whose both radii are proportional to $N^{1/3}$. At 
large $N$, the Maldacena LEEA is given by a $(2,0)$ superconformally invariant gauge field theory in
six dimensions, which is supposed to be dual to M-theory compactified on $AdS_7\times S^4$ 
(\cite{mal}). Down to four spacetime dimensions, Maldacena (1997) considered $N$ D-3-branes at large
$N$ instead, and he argued that the N=4 super-Yang-Mills theory in the t'Hooft limit has to be dual 
to the IIB superstring theory compactified on $AdS_5\times S^5$. The four-dimensional N=4 
super-Yang-Mills theory is supposed to live on the boundary of the $AdS_5$-space, with the 
correspondence
\begin{equation}
Ng^{2}_{\rm YM}\sim (\a')^{-2}R^4_{\rm AdS}~,\qquad g^2\low{\rm YM}\sim g\low{\rm string} ~.
\label{ac}
\end{equation}
Note that the t'Hooft limit of large $Ng^{2}_{\rm YM}$ is equivalent to $\abs{\x}\to 0$ in our
approach because of eq.~({\ref{coup}). 

To conclude, the conformal (t'Hooft-Maldacena) LEEA limit described by the N=4 super-Yang-Mills 
theory can be deduced from the hypermultiplet LEEA near the singularity, after taking into account 
the dynamical generation of the massless hypermultiplets in the adjoint representation of the
gauge group, and of the non-abelian N=2 vector gauge multiplets at a non-vanishing N=2 
central charge. Our approach can be generalized to any simply-laced gauge group.

\section*{Acknowledgements}

I would like to thank Dieter L\"ust and J\"urgen Schulze for useful discussions.

\end{document}